\documentclass[conference]{IEEEtran}
\usepackage{geometry}
\usepackage{booktabs} 
\usepackage{graphicx}
\usepackage{color}
\usepackage[linesnumbered,ruled]{algorithm2e}
\usepackage{ amssymb }
\usepackage{ mathrsfs }
\usepackage{float}
\usepackage{amsmath}
\usepackage{multirow}
\usepackage{transparent}
\usepackage{subcaption}
\usepackage{blindtext}
\usepackage{booktabs}
\usepackage[table]{xcolor}
\usepackage{enumitem}
\makeatletter
\def\ps@IEEEtitlepagestyle{%
  \def\@oddfoot{\mycopyrightnotice}%
  \def\@evenfoot{}%
}
\def\mycopyrightnotice{%
  {\footnotesize The copyright belongs to me!\hfill}
  \gdef\mycopyrightnotice{}
}
\newcommand\blfootnote[1]{%
  \begingroup
  \renewcommand\thefootnote{}\footnote{#1}%
  \addtocounter{footnote}{-1}%
  \endgroup
}

\begin{document}
\date{}

\title{\Large\textbf{Characterizing and Optimizing EDA Flows for the Cloud}}	

\makeatletter
\def\ps@IEEEtitlepagestyle{%
  \def\@oddfoot{\mycopyrightnotice}%
  \def\@evenfoot{}%
}
\makeatother
\def\mycopyrightnotice{%
  \begin{minipage}{\textwidth}
    \footnotesize
  \end{minipage}
  \gdef\mycopyrightnotice{}
}

\author{\normalsize
	\begin{tabular}[t]{c@{\extracolsep{1em}}c@{\extracolsep{1em}}c@{\extracolsep{1em}}c}
		\large Abdelrahman Hosny& \large Sherief Reda \\
		Department of Computer Science & School of Engineering \\
		Brown University  & Brown University \\
		abdelrahman\_hosny@brown.edu & sherief\_reda@brown.edu\\
\end{tabular}}

\maketitle

{\small\textbf{Abstract---
Cloud computing accelerates design space exploration in logic synthesis, and parameter tuning in physical design.
However, deploying EDA jobs on the cloud requires EDA teams to deeply understand the characteristics of their jobs in cloud environments. Unfortunately, there has been little to no public information on these characteristics. Thus, in this paper, we formulate the problem of migrating EDA jobs to the cloud. First, we characterize the performance of four main EDA applications, namely: synthesis, placement, routing and static timing analysis. We show that different EDA jobs require different machine configurations. Second, using observations from our characterization, we propose a novel model based on Graph Convolutional Networks to predict the total runtime of a given application on different machine configurations. Our model achieves a prediction accuracy of 87\%. Third, we develop a new formulation for optimizing cloud deployments in order to reduce deployment costs while meeting deadline constraints. We present a pseudo-polynomial optimal solution using a multi-choice knapsack mapping that reduces costs by 35.29\%. 
}}

\blfootnote{This work is partially supported by NSF grant 1814920.}
\vspace{-0.1in}
\section{Introduction}
\label{sec:intro}

EDA (Electronic Design Automation) tools expose hundred of parameters to tune for front-end and back-end design. 
Therefore, design space exploration and efficient physical implementation have been increasingly challenging and require a massive amount of compute to achieve acceptable Quality of Results (QoR).
In the recent years, there has been a growing trend among EDA teams to utilize elastic compute environments (i.e. cloud) to gain near-instant access to compute resources \cite{intel-cloud}.
Migrating EDA jobs to the cloud has helped teams meet the demands of their tapeout schedule, hence reducing the time to market \cite{sehgal2016eda-ready-cloud}.
For example, horizontal scaling by launching more compute servers allows EDA teams to complete a highly-parallelizable compute job such as simulation in less time.
In addition, EDA teams have the flexibility to choose the configuration of hardware that meets their needs for the exact pending job and only for the time needed to complete it.

However, migrating EDA jobs to the cloud is not a straightforward path, especially for teams with little or no experience managing elastic resources. 
For example, design teams need to choose the right machine configuration that achieves the best performance for their job.
While simulation and verification are known to be embarrassingly parallel (i.e. directly benefiting from the scale of the cloud), the compute requirements for the synthesis and physical design stages are not well-studied, especially in multi-tenancy environments.
Furthermore, reducing deployment costs while meeting the tapeout schedule is a challenge, especially in teams with limited budget.

\begin{figure}[t]
    \centering
	\includegraphics[scale=0.5]{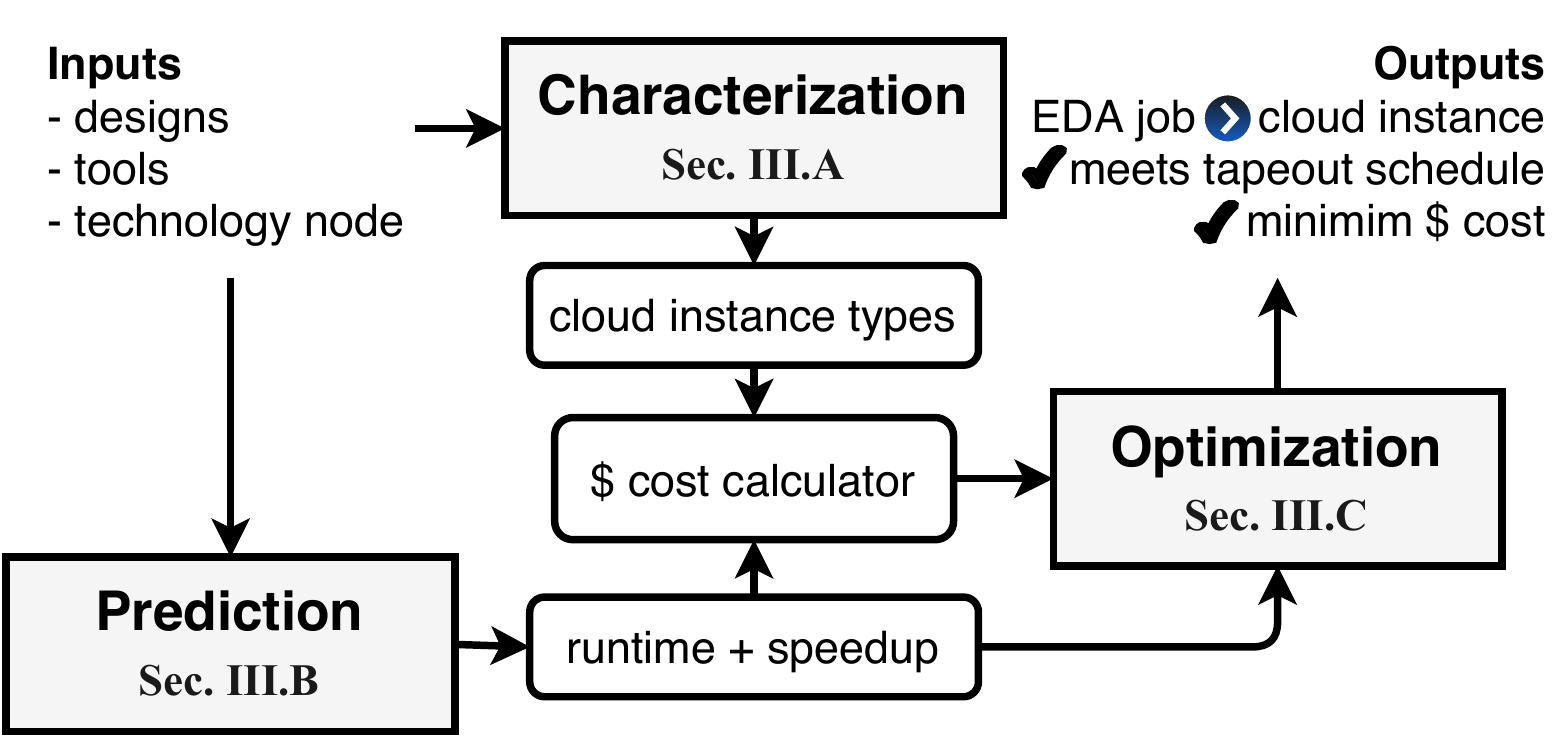}
    \caption{Workflow of optimizing EDA cloud deployments}
    \vspace{-0.25in}
    \label{fig:overview}
\end{figure}

Our main contributions are:

\begin{enumerate}[leftmargin=*]
    \item We characterize the performance of four EDA key applications (synthesis, placement, routing, and static timing analysis) under different machine configurations. Using our observations, we present recommendations for the configurations of cloud instances to provision for each application.
    \item We propose a novel model based on Graph Convolutional Networks (GCNs) that predicts the total runtime a given job would take using certain machine configurations. Our model achieves a high prediction accuracy of 87\%.
    \item With recommended machine configurations and runtime predictions for each application, we reduce cloud deployment costs subject to deadline constraints by mapping the problem to the classical multi-choice knapsack problem (NP-hard). Our implementation recommends optimal machine configurations that minimizes the total cloud deployment cost by an average of 35.29\%. 
\end{enumerate}

Next, we give a brief background in Section \ref{sec:background}. 
In Section \ref{sec:opt}, we formulate the problem and discuss our proposed framework. 
In Section \ref{sec:experiment}, we present our experimental results.

\begin{figure*}[t!]
    \begin{subfigure}[]{0.25\textwidth}
        \centering
        \vspace{-0.2in}
        \includegraphics[scale=0.29]{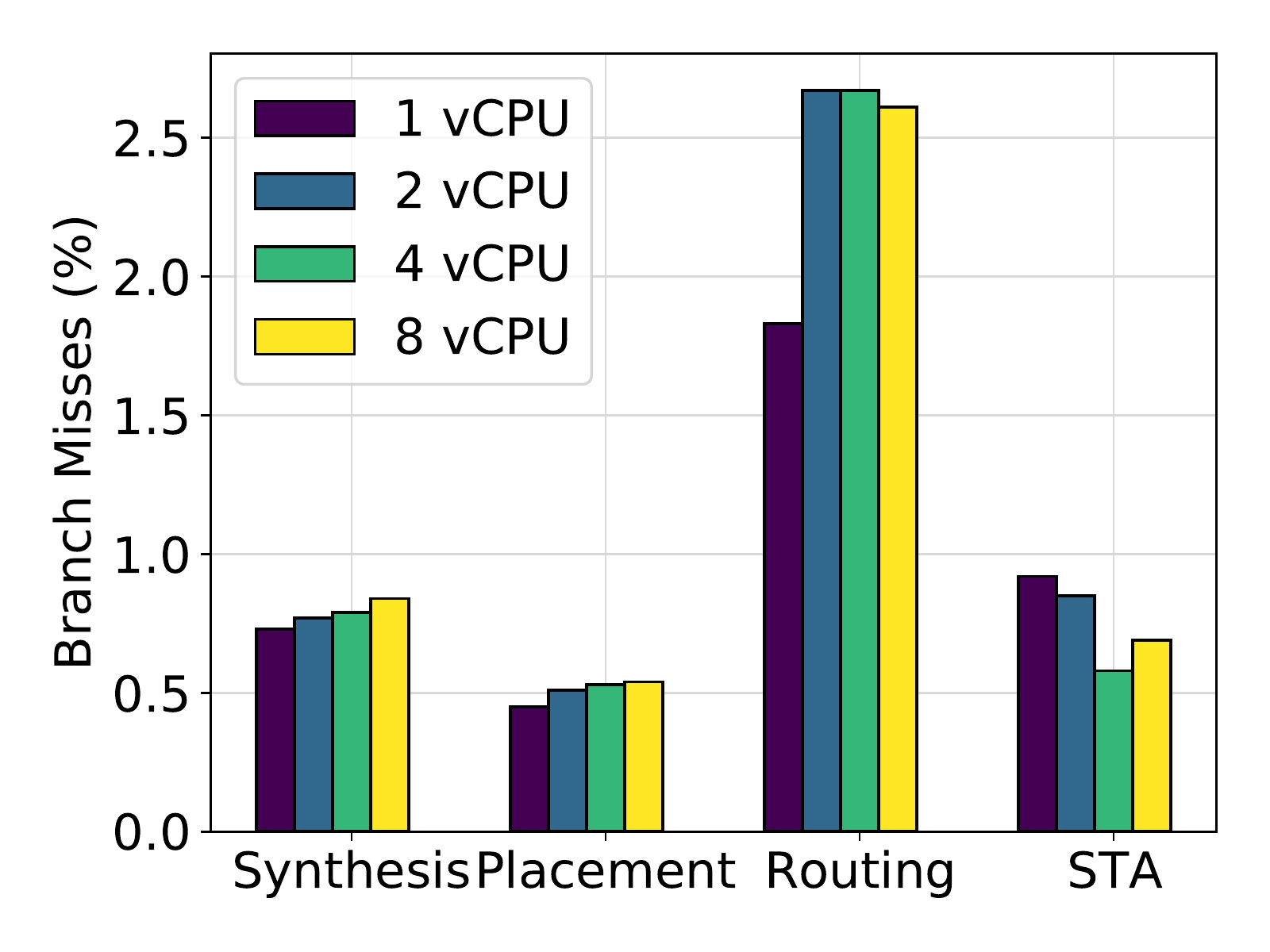}
        \vspace{-0.25in}
        \caption{Branch Misses}
    \end{subfigure}%
    \begin{subfigure}[]{0.25\textwidth}
        \centering
        \vspace{-0.2in}
        \includegraphics[scale=0.29]{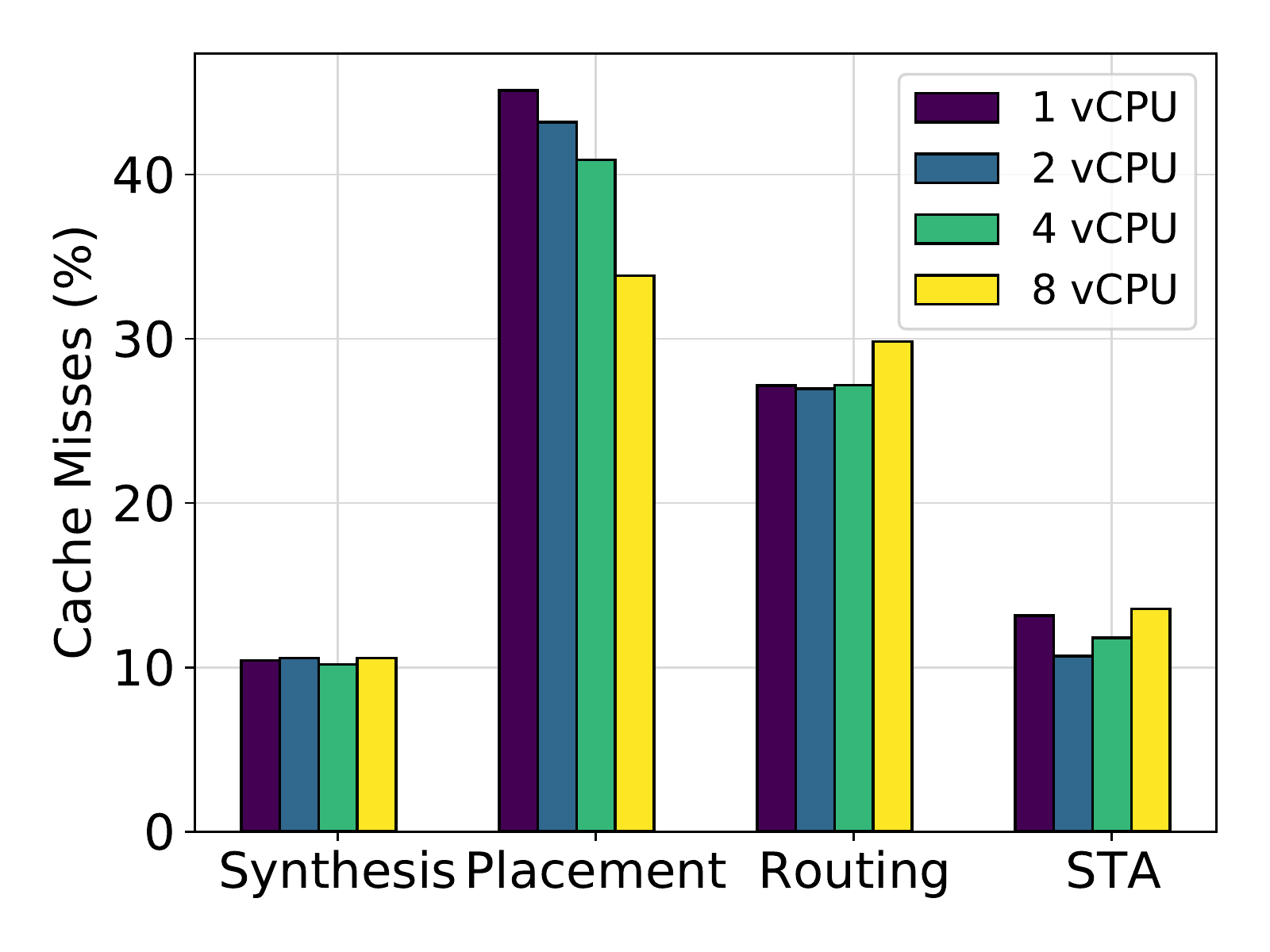}
        \vspace{-0.25in}
        \caption{Cache Misses}
    \end{subfigure}%
    \begin{subfigure}[]{0.25\textwidth}
        \centering
        \vspace{-0.2in}
        \includegraphics[scale=0.29]{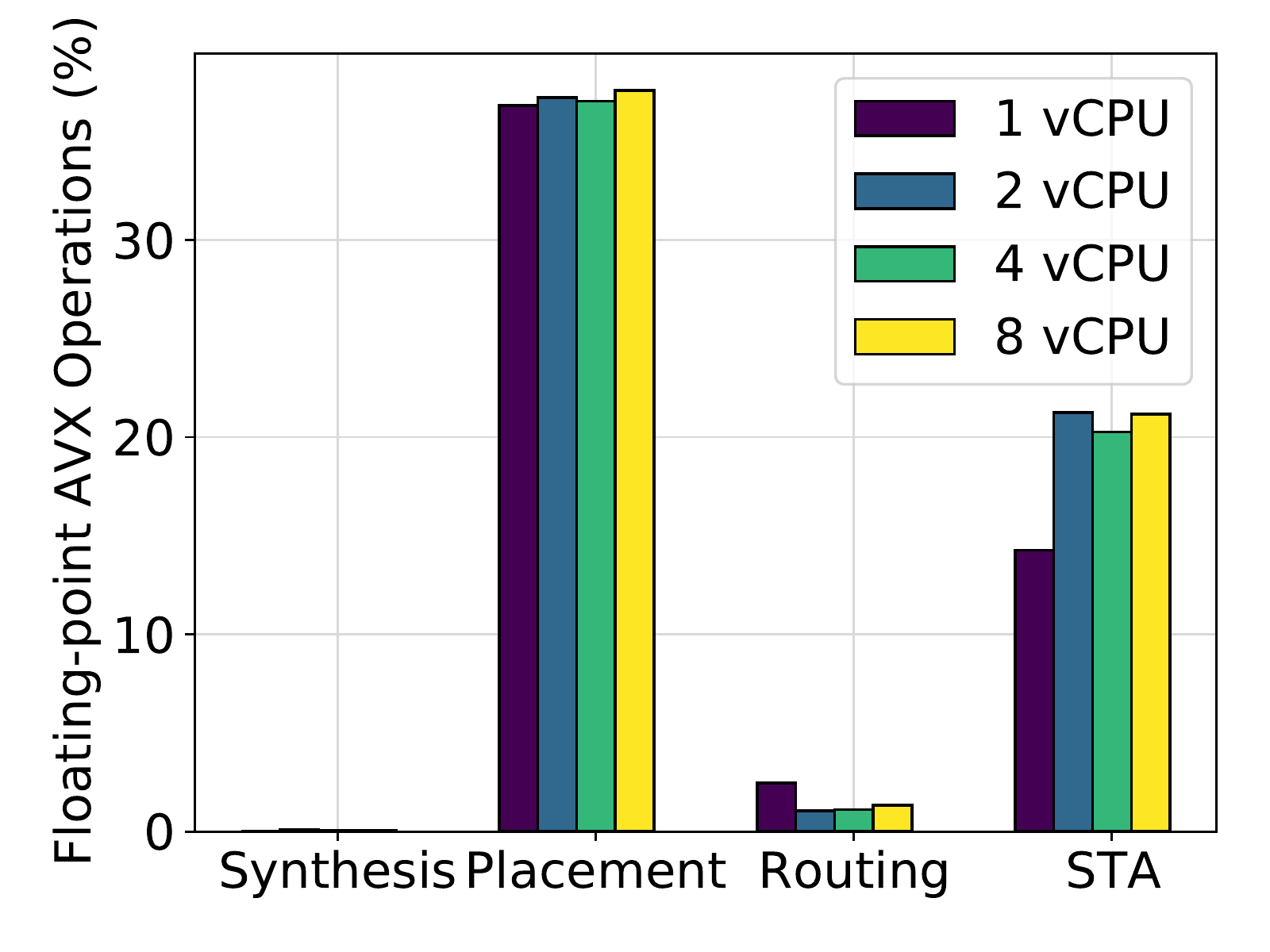}
        \vspace{-0.25in}
        \caption{Floating-point Operations}
    \end{subfigure}%
    \begin{subfigure}[]{0.25\textwidth}
        \centering
        \vspace{-0.2in}
        \includegraphics[scale=0.29]{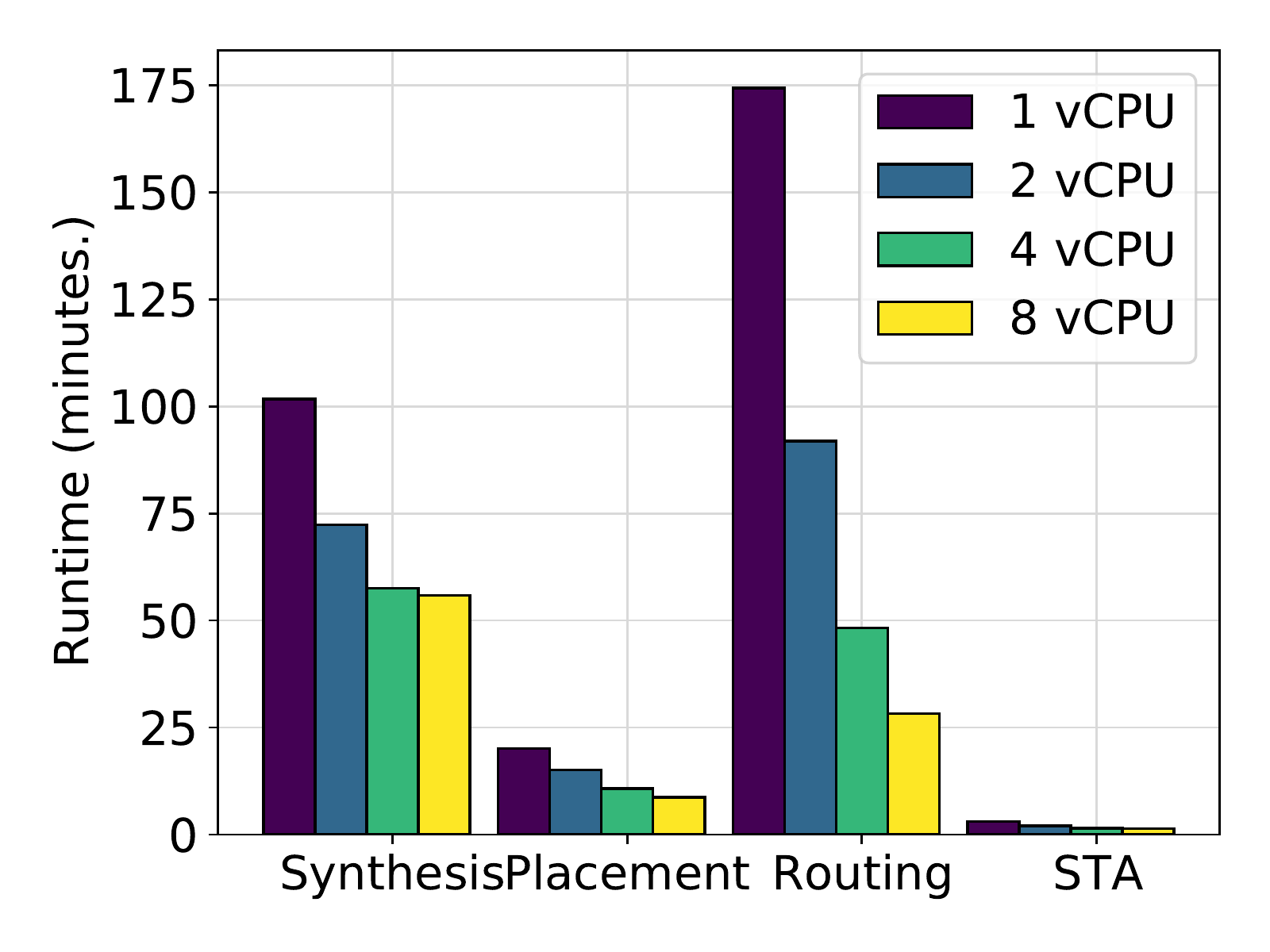}
        \vspace{-0.25in}
        \caption{Total Runtime}
    \end{subfigure}
    \caption{Performance characterization of four representative EDA jobs} 
    \vspace{-0.2in}
    \label{fig:char-compute}
    
\end{figure*}

\section{Background}
\label{sec:background}

Cloud computing refers to the elastic compute resources that can be provisioned, scaled up or shutdown on demand. 
Cloud providers virtualize their physical infrastructure to share processing time, memory, storage and network bandwidth among many users (known as tenants). 
In order to achieve this virtualization, cloud vendors use a specialized software called the \textit{Hypervisor}. The Hypervisor isolates each tenant's resources in a Virtual Machine (VM) that is accessible only by its owner.
In a standard cloud offering, VMs are sold in units of: (i) vCPU: a virtual CPU is seen as a single CPU thread, (ii) Memory: a fixed number of memory pages is solely reserved for the use of a VM and is expressed as the total memory size reserved, and (iii) Storage: the size and type of the underlying storage device partition mounted on the VM.



\section{Optimizing EDA Cloud Deployments}
\label{sec:opt}

\textbf{Problem Definition.}
A fundamental question that faces EDA teams when migrating their EDA jobs to the cloud is: what configurations of VMs should be provisioned for each job? And how can the job completion time be reduced while minimizing the cost?

In order to answer these questions, Figure \ref{fig:overview} draws our workflow that we propose in this paper. Specifically, we introduce the following problems:

\noindent\textbf{Problem 1.}
What is the right VM configuration for a given EDA job?
To address this problem, we characterize four main EDA applications, namely: synthesis, placement, routing and static timing analysis. 
We focus on characteristics that are intrinsic to the EDA job which affect the completion time.

\noindent\textbf{Problem 2:} Given a design (in RTL or Netlist), predict the runtime for a given job (e.g. synthesis or routing) when using 1, 2, 4 and 8 vCPUs. Our proposed prediction model, learns internal graph features of the design that affect the total runtime of a given job on different machine sizes.

\noindent\textbf{Problem 3:} Given the runtime for each job under 1, 2, 4 and 8 vCPUs, as well as a deadline constraint, select a machine size for each job such that the deadline is met and the total cost is minimized.
We address this problem using a mapping to the multi-choice knapsack problem and implement an optimal solution using dynamic programming.

\subsection{EDA Flow Characterization}
\label{sec:char}

To address Problem 1, we characterized the four jobs using a major commercial EDA flow and a SPARC core design from OpenPiton design benchmark \cite{balkind2016openpiton} using a 14nm technology node.
We then collected the execution data from the system's hardware performance counters for further analysis.
In order to simulate a multi-tenancy environment, we used Linux Control Groups on a machine with a 14-core Intel Xeon E5-2680 processor running at 3.3GHz, and 128GB DDR4 memory.
We used the Linux perf utility to instrument the hardware
performance counters.

\noindent\textbf{Branch Prediction.} Figure \ref{fig:char-compute}-a summarizes our findings from the characterization experiments.
First, we observe that routing has a higher percentage of branch misses.
We attribute this value to the nature of the routing algorithms, where there can be few trials before a net is successfully routed with no design rule violations.
In particular, graph search algorithms in the routing step encompass a large portion of conditional statements that cannot be avoided.
Rip-up and reroute techniques also contribute to halting the continuous execution of the routing algorithms.

\noindent\textbf{Memory Access Patterns.}
In Figure \ref{fig:char-compute}-b, we observe that placement and routing have significantly higher cache misses than synthesis and STA.
Placement has a 45.11\% cache misses rate when using 1 vCPU and 33.84\% when using 8 vCPUs, while routing has 27.15\% and 29.84\%  cache misses rate using 1 and 8 vCPUs respectively.
We attribute this higher miss rate to the nature of the analytical component in the placement engine that tries to optimize the wirelength across all the chip instances using convex optimization methods. 
This needs access to large vectors to calculate the gradients, hence benefiting from the more cache available with more vCPUs.

\noindent\textbf{Floating-point Operations.} 
In Figure \ref{fig:char-compute}-c, we observe that the placement job requires more floating-point operations that run on Advanced Vector Extensions (AVX) hardware.
This can be attributed to the analytical engine that tries to optimize the wire length across all the chip area using convex optimization methods.
This involves calculating gradients which relies on floating-point operations.
The STA job comes next in its percentage usage of the AVX hardware.
This is consistent with the nature of STA algorithms where calculating slacks involves graph traversal from inputs to outputs, with access to floating-point values in the technology library.

\begin{figure}[t]
    \centering
    \vspace{-0.1in}
	\includegraphics[scale=0.4]{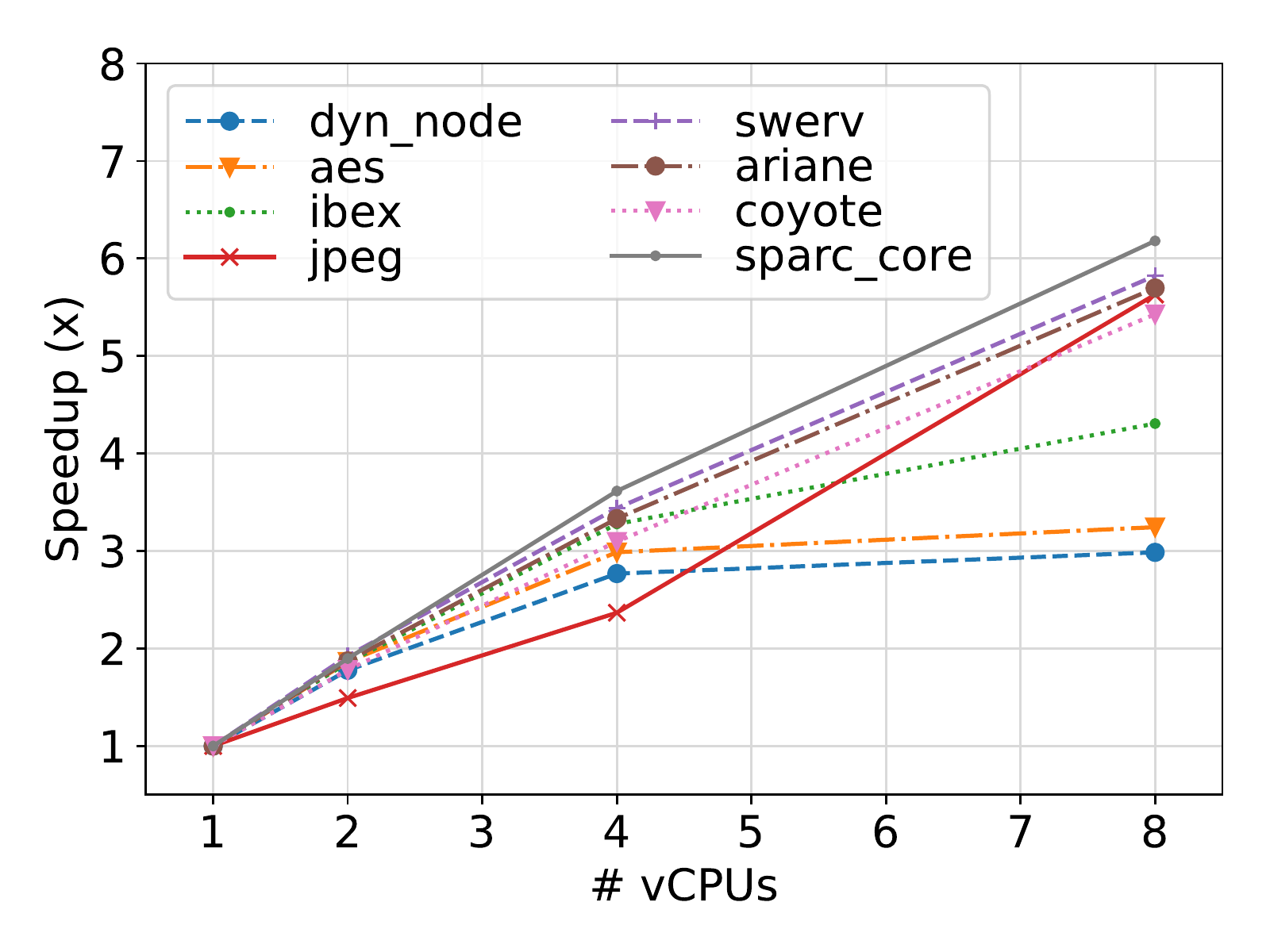}
	\vspace{-0.15in}
    \caption{Routing speedup for different designs. \textit{dyn\_node} is the smallest and \textit{sparc\_core} is the largest (\#instances).}
    \vspace{-0.3in}
    \label{fig:speedup}
\end{figure}

\noindent\textbf{Scalability and Speedup.}
In Figure \ref{fig:char-compute}-d, we observe that the routing job scales well with more \#vCPUs.
This is consistent with the nature of the routing job, where nets in independent grid cells can be routed in parallel with no conflict, as opposed to synthesis, placement and STA where internal algorithms have inherent dependencies.
Further analysis of the routing job, Figure \ref{fig:speedup} plots the speedups achieved on different designs of different characteristics and sizes.
It shows that adding more vCPUs does not eminently scale the routing job in all designs.
Smaller designs (such as dynamic\_node and aes) have almost equal speedups for 4 and 8 vCPUs.
This means that the provisioned vCPUs might not offer the expected benefit from the cloud scale, and that there is an opportunity to achieve the same outcome in nearly the same time with less resources.

\noindent\textbf{Main Takeaways.}
From the point of view of EDA teams running their EDA applications on the cloud, we summarize our main recommendations:
\begin{enumerate}
    \item Synthesis and STA jobs perform well on general-purpose VM instances with a balance between computations and memory access. Placement and routing require VM instances with higher memory-to-core ratio, with routing demanding more available L1 and LLC cache.
    \item Placement jobs should be run on a compute instance with an underlying processor that supports Advanced Vector Extensions (AVX). STA jobs would also benefit from AVX hardware.
    \item On large designs, routing jobs scale well with the number of vCPUs allocated. However, on small designs, speedup is capped at a certain point.
\end{enumerate}

\noindent These observations motivate our work in the next section.


\subsection{Runtime Prediction}
\label{sec:prediction}

\begin{figure}[t!]
	\includegraphics[scale=0.45]{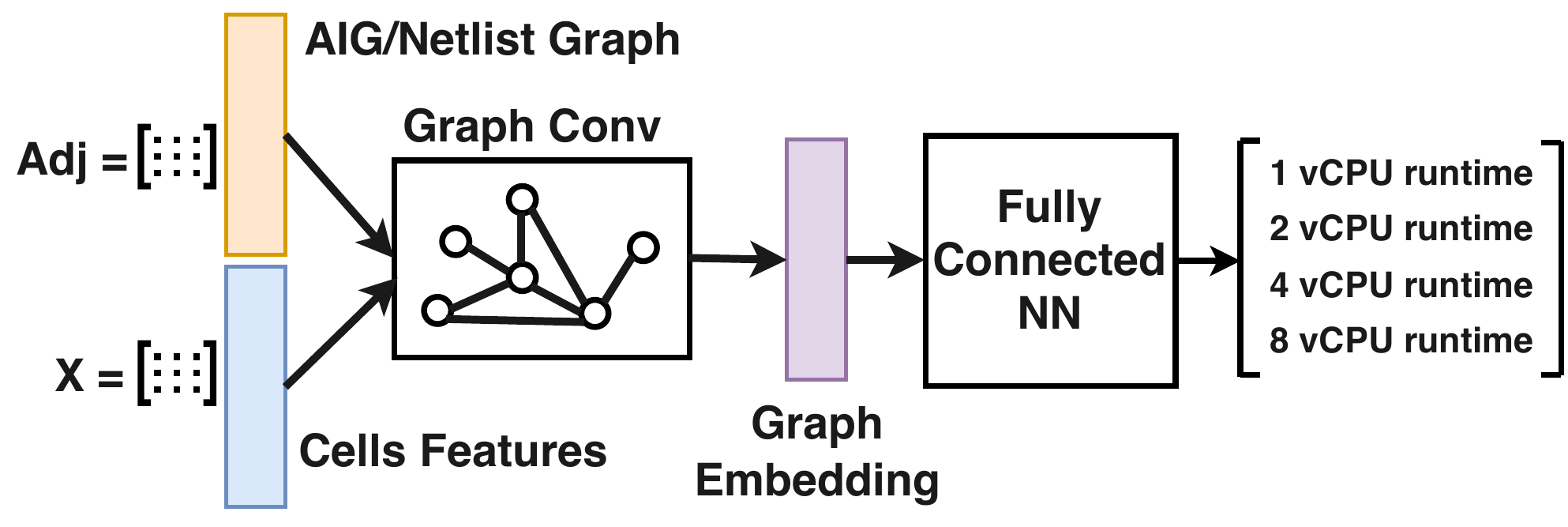}
    \caption{Our proposed runtime prediction model}
    \vspace{-0.25in}
    \label{fig:proposed-model}
\end{figure}

To address Problem 2, we state that the runtime of chip design tasks depends on a number of factors such as the design itself, the tools used, the technology node, the parameters used to instruct the tools and the VM configuration.
Without losing generality, when using the same tools, technology node, default parameters and VM configuration, the runtime of a certain job depends on the complexity of the design itself.
Figure \ref{fig:proposed-model} shows the architecture of our model.
The model takes as input the design in RTL or netlist, and performs an embedding operation using Graph Convolutional Networks \cite{kipf2016semi}.
After that, a fully-connected neural layer transforms the embedding into predictions for the runtime under different machine sizes (i.e. \#vCPUs).
This model is trained for each application separately.
Using the predicted runtimes, we can calculate the speedup gains from using 2, 4 or 8 vCPUs as compared to using only 1 vCPU.

\noindent\textbf{Processing Input Design.}
When building a model to predict synthesis runtime, the input is usually in RTL, which is not a graph.
However, synthesis tools map the RTL into an intermediate representation such as And-Inverter Graphs (AIG) before synthesizing and mapping to a technology library.
Therefore, our model can operate on the AIG representation of the design. 
The AIG is a Directed Acyclic Graph (DAG), which means it preserves edge directions for the GCN.
On the other hand, when building a prediction model for the placement and routing, the input is expected to be a netlist.
In order to operate on the given netlist, we convert cells and I/O pins into graph nodes. In addition, we convert each net into a set of directed edges using the well-known star model, where there is one edge from the driving cell (or the input pin) towards each of the sinks (or the output pins).

\noindent\textbf{Graph Convolutions.}
In Graph Convolutional Networks (GCNs), the key idea is to generate node embeddings based on local neighborhoods. 
The first layer's embedding of a node represents its input feature vector, $x_i$. Nodes aggregate information from their neighbors in each convolutional layer. 
This aggregation is followed by an activation function, such as \textit{ReLU}, and a pooling operation, such as \textit{sum}-pooling. With that in-place, every layer is written as a non-linear function:

\vspace{-0.1in}
\begin{equation}
\mathcal{H}^{(l+1)} = f(\mathcal{H}^{(l)}, \mathcal{A}) \, ,
\label{eq:1}
\vspace{-0.05in}
\end{equation}

\noindent where $\mathcal{H}^{(l)}$ represents the activation at layer $l$, and $\mathcal{H}^{(0)}$ is the input feature matrix, $\mathcal{X}$. $\mathcal{A}$ is the adjacency matrix. Looking at the embedding of each node, we can elaborate on Equation \ref{eq:1} as follows:

\vspace{-0.18in}
\begin{equation*}
    h_v^k = \sigma\ (W_k\ \sum\nolimits_{_u \in N(v)} \frac{h_u^{k-1}}{|N(v)|}\ +\ B_k h_v^{k-1}) \quad \forall k \in \{1,...,K\}
    \label{eq:2}
\end{equation*}

\noindent where, $h_v^k$ is a vector that represents $k^{th}$-layer embedding of node $v$. $N(v)$ represents the neighbors of node $v$. $W_k$ and $B_k$ are the trainable matrices which are shared with all nodes of the graph, and $\sigma$ is the activation function. After $K$-layers of neighborhood aggregation, we get output embeddings for each node that can be fed into a loss function. We can then run stochastic gradient descent to learn $W_k$ and $B_k$.

\noindent\textbf{Model Design.}
We used 2 GCN layers with 256 and 128 hidden units each, followed by 1 fully connected layer with 128 units.
The model is trained for 200 epochs using Mean Square Error (MSE) as a loss function and Adam as the optimizer (lr=1e-4).
The loss function calculates the combined prediction error for all four runtimes (i.e. 1, 2, 4 and 8 vCPUs).

\begin{figure*}[t!]
\begin{minipage}{0.25\textwidth}
\centering
\includegraphics[width=1.8in,height=1.3in]{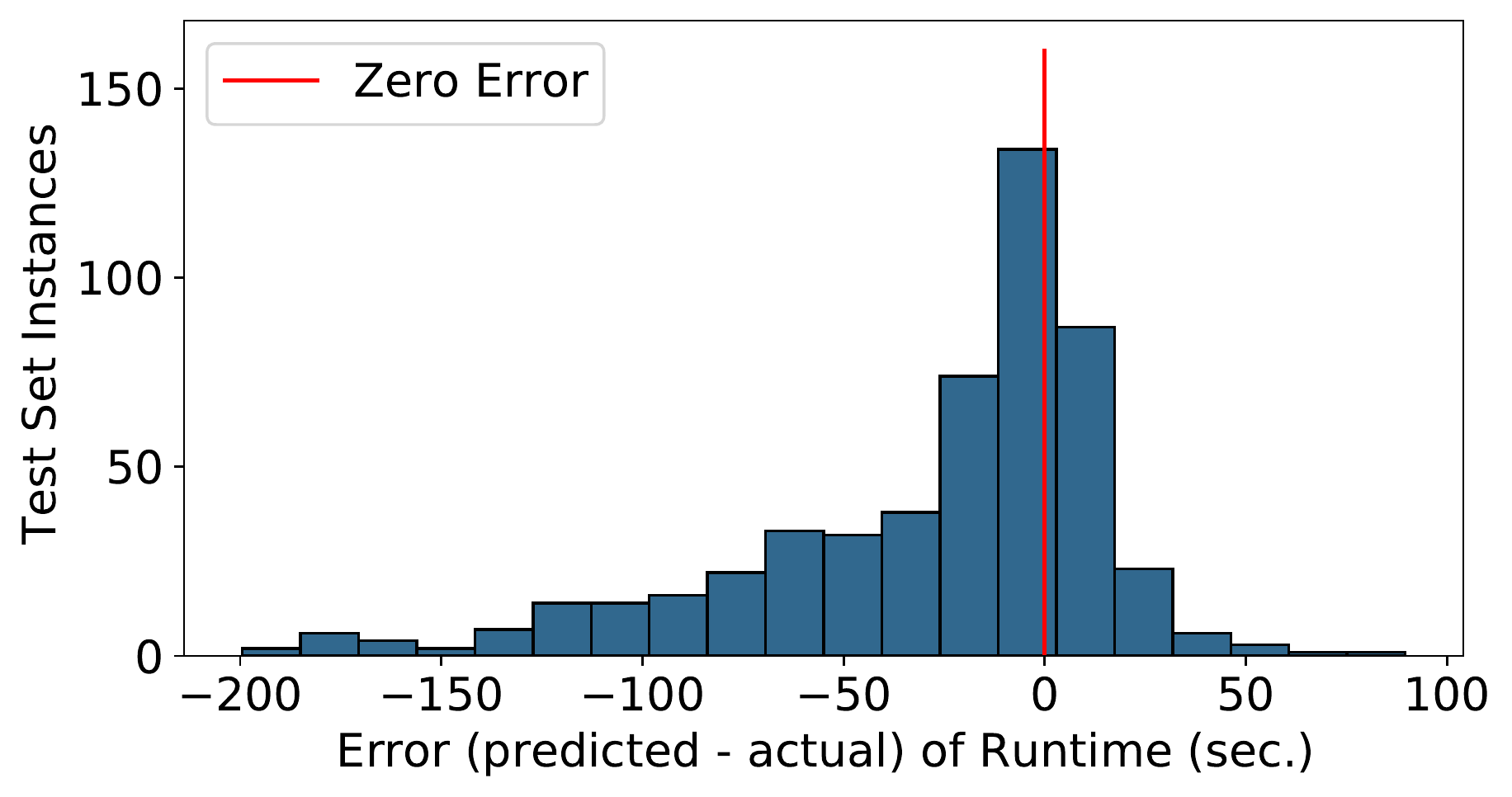}
\caption{\centering Runtime prediction errors. Avg. Error: 13\%.} \label{fig:predictions}
\end{minipage}
\begin{minipage}{0.75\textwidth}
\captionsetup{type=table} 
\scriptsize
\begin{tabular}{|l|l|c|c|c|c|c|c|c|c|c|c|c|c|c|c|c|c|c|c|}
\hline
\multicolumn{2}{|l|}{\textbf{Task}} & \multicolumn{4}{c|}{\begin{tabular}[c]{@{}c@{}}\textbf{Synthesis}\\ general-purpose VM\end{tabular}} & \multicolumn{4}{c|}{\begin{tabular}[c]{@{}c@{}}\textbf{Placement}\\ memory-optimized VM\end{tabular}} & \multicolumn{4}{c|}{\begin{tabular}[c]{@{}c@{}}\textbf{Routing}\\ memory-optimized VM\end{tabular}} & \multicolumn{4}{c|}{\begin{tabular}[c]{@{}c@{}}\textbf{STA}\\ general-purpose VM\end{tabular}} & \multirow{4}{*}{\begin{tabular}[c]{@{}c@{}}\rotatebox[origin=c]{270}{\textbf{Total Runtime}}\end{tabular}} & \multirow{4}{*}{\begin{tabular}[c]{@{}c@{}}\rotatebox[origin=c]{270}{\textbf{Min Cost} (\$)}\end{tabular}} \\ \cline{1-18}
\multicolumn{2}{|l|}{\textbf{vCPUs}} & 1 & 2 & 4 & 8 & 1 & 2 & 4 & 8 & 1 & 2 & 4 & 8 & 1 & 2 & 4 & 8 &  &  \\ \cline{1-18}
\multicolumn{2}{|l|}{\textbf{Runtime (sec.)}} & \multicolumn{1}{r|}{6100} & \multicolumn{1}{r|}{4342} & \multicolumn{1}{r|}{3449} & \multicolumn{1}{r|}{3352} & \multicolumn{1}{r|}{1206} & \multicolumn{1}{r|}{905} & \multicolumn{1}{r|}{644} & \multicolumn{1}{r|}{519} & \multicolumn{1}{r|}{10461} & \multicolumn{1}{r|}{5514} & \multicolumn{1}{r|}{2894} & \multicolumn{1}{r|}{1692} & \multicolumn{1}{r|}{183} & \multicolumn{1}{r|}{119} & \multicolumn{1}{r|}{90} & \multicolumn{1}{r|}{82} &  &  \\ \cline{1-18}
\multicolumn{2}{|l|}{\textbf{Cost (\$)}} & \multicolumn{1}{r|}{0.16} & \multicolumn{1}{r|}{0.15} & \multicolumn{1}{r|}{0.19} & \multicolumn{1}{r|}{0.37} & \multicolumn{1}{r|}{0.04} & \multicolumn{1}{r|}{0.04} & \multicolumn{1}{r|}{0.05} & \multicolumn{1}{r|}{0.08} & \multicolumn{1}{r|}{0.32} & \multicolumn{1}{r|}{0.25} & \multicolumn{1}{r|}{0.21} & \multicolumn{1}{r|}{0.25} & \multicolumn{1}{r|}{0.02} & \multicolumn{1}{r|}{0.01} & \multicolumn{1}{r|}{0.02} & \multicolumn{1}{r|}{0.05} &  &  \\ \hline
\multirow{4}{*}{\begin{tabular}[c]{@{}l@{}}\textbf{Total} \\ \textbf{Runtime}\\ \textbf{Constraint}\end{tabular}} & \multicolumn{1}{r|}{10000} &  & x &  &  & x &  &  &  &  &  & x &  &  & x &  &  & \multicolumn{1}{r|}{8561} & \multicolumn{1}{r|}{0.41} \\ \cline{2-20} 
 & \multicolumn{1}{r|}{6000} &  &  & x &  &  &  & x &  &  &  &  & x &  & x &  &  & \multicolumn{1}{r|}{5904} & \multicolumn{1}{r|}{0.50} \\ \cline{2-20} 
 & \multicolumn{1}{r|}{5645} &  &  &  & x &  &  &  & x &  &  &  & x &  &  &  & x & \multicolumn{1}{r|}{5645} & \multicolumn{1}{r|}{0.75} \\ \cline{2-20} 
 & \multicolumn{1}{r|}{5000} &  &  &  &  &  &  &  &  &  &  &  &  &  &  &  &  & \multicolumn{1}{r|}{NA} & \multicolumn{1}{r|}{NA} \\ \hline
\end{tabular}
\caption{Minimizing  total  cloud  deployment  cost  subject to a time constraint. The mark (x) denotes the recommended machine configuration. NA denotes Not Achievable.}
\label{tab:opt-results}
\end{minipage}
\vspace{-0.25in}
\end{figure*}

\subsection{Cloud Deployment Optimization}
\label{sec:knapsack}

Given the runtime estimates, we now address Problem 3.
Our proposed solution maps the problem to the multi-choice knapsack problem (MCKP) \cite{Kellerer2004}. 
Using our predictions calculated in the previous section, each job can be run on a different machine configuration (i.e. \#vCPUs), where each configuration completes the job in $t$ time and costs $p$ in total.

\noindent\textbf{Formulation.}
Let $z_l(C)$ be an optimal solution defined on $l$ applications and with total runtime constraint $C$:

\vspace{-0.1in}
\begin{equation}
    z_l(C) := \max \sum_{i=1}^{l} \sum_{j=1}^{N_i} s_{ij} \frac{1}{p_{ij}}
\vspace{-0.1in}
\end{equation}
\vspace{-0.2in}
\noindent such that,
\begin{align*}
    & \sum_{i=1}^{l} \sum_{j=1}^{N_i} s_{ij} t_{ij}  \leq C, \\
    & \sum\nolimits_{j \in N_i} s_{ij} = 1, i = 1,....,l, \\
    & s_{ij} \in \{0,1\}, i=1,....,l, j \in N_i
\end{align*}

\noindent where $s_{ij} \in \{0, 1\}$ denotes whether we select VM configuration $j$ for stage $i$ or not, and $N_i$ denotes the number of configurations in a given stage.
$t_{ij}$ denotes the runtime of stage $i$ when using $j$'s configuration, which is obtained from the runtime predictions. 
Similarly, $p_{ij}$ denotes the cost of running stage $i$ when using $j$'s configuration, which is obtained from the pricing table of the selected cloud vendor.
We assume that $z_l(C) := -\infty$ if no solution exists (i.e. the total runtime is not sufficient to complete all the stages using the fastest machine configuration).

To solve (2), we implemented a pseudopolynomial solution through dynamic programming utilizing Dudzinski and Walukiewicz approach \cite{dudzinski1987exact}:

\vspace{-0.2in}
\begin{align*}
    z_l(C) = \max \left\{
                \begin{array}{ll}
                  z_{l-1}(C-t_{l1}) + 1 / p_{l1}\ \  & \mbox{if}\ 0 \leq C - t_{l1},\\
                  z_{l-1}(C-t_{l2}) + 1 / p_{l2}\ \  & \mbox{if}\ 0 \leq C - t_{l2},\\
                  :\\
                  z_{l-1}(C-t_{l_{n_l}}) + 1 / p_{l_{n_l}}\ \  & \mbox{if}\ 0 \leq C - t_{l_{n_l}}
                \end{array}
              \right.
\end{align*}

This implementation provides an optimal solution provided that the runtime values are rounded to the nearest integer (second).
This is an assumption that we can safely make in our case since cloud machines are billed per second (no fractions).

\section{Experimental Results}
\label{sec:experiment}
We demonstrate our predictions using GF 14nm technology node and a major commercial EDA flow.

\noindent\textbf{Dataset.} We use 18 representative benchmarks of different sizes and structures from EPFL benchmark suite \cite{amaru2015epfl} and OpenCores \cite{opencores}. 
We synthesize each benchmark applying different logic optimizations to generate different netlists. 
The motivation is to challenge the GCN with netlists that have different physical structures, but perform the same logic function. 
We have a total of 330 unique netlists, with 2,640 data points (runtimes) for different machine configurations.
These designs range from a few hundred instances to ~200k instances.
We divide the dataset into training and test groups with 80\% and 20\% respectively, where netlists of the test set belong to unseen designs in the training set.

\noindent\textbf{Prediction Accuracy.}
Due to lack of space, we show a histogram of model prediction errors for placement and routing in Figure \ref{fig:predictions}. 
Runtime predictions given a netlist (placement, routing, STA) achieves an average error of 13\%.
On AIGs (synthesis), the runtime prediction has an average error of 5\%.

\begin{figure}[t]
    \centering
    \includegraphics[scale=0.38]{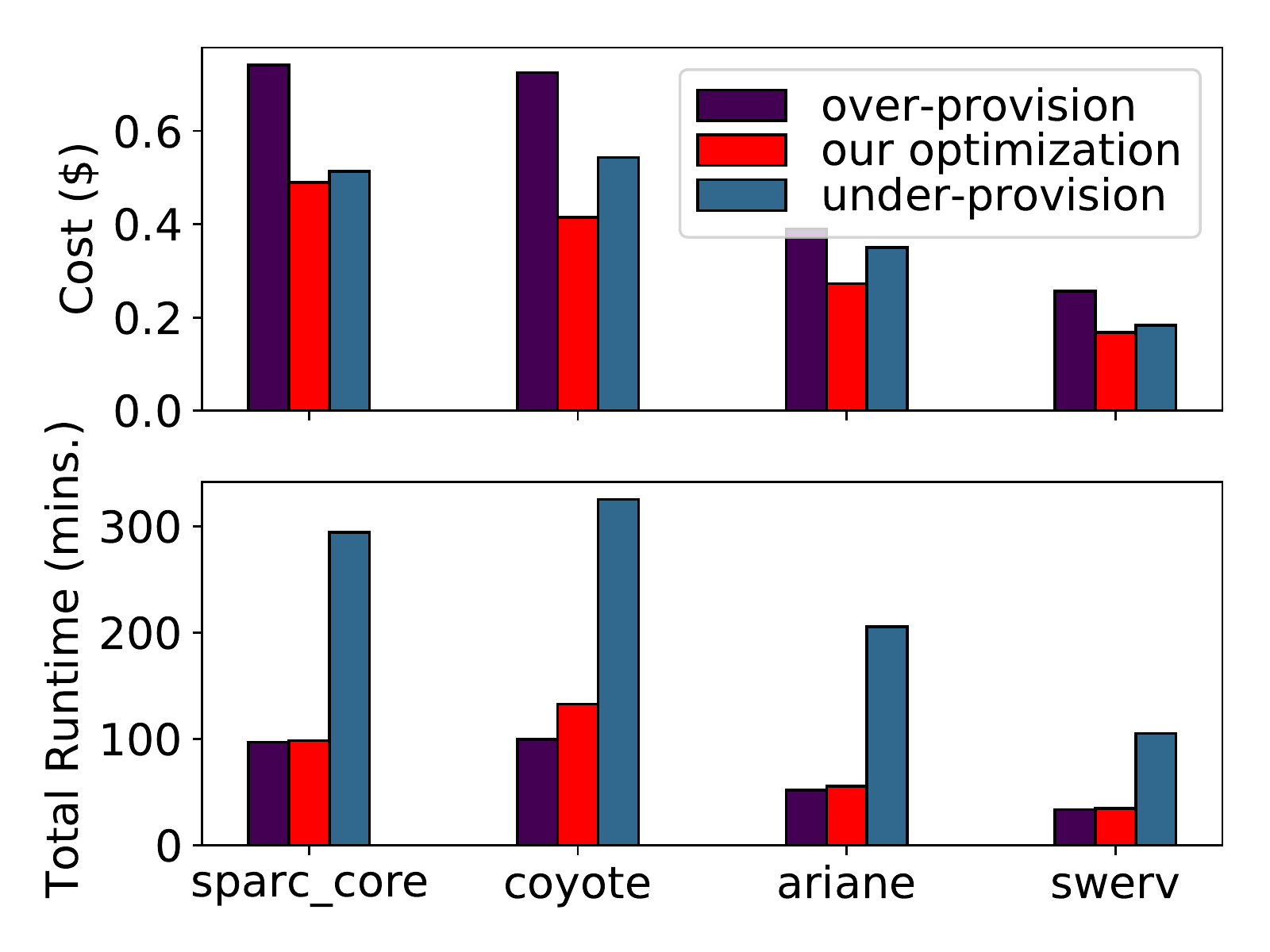}
    \vspace{-0.1in}
    \caption{Cost savings from running our muli-choice knapsack optimization algorithm. Over-provisioning runs all stages on 8 vCPUs. Under-provisioning runs all stages on 1 vCPUs.}
    \vspace{-0.25in}
    \label{fig:cost}
\end{figure}

\noindent\textbf{Optimization Results.}
Referring to Figure \ref{fig:overview}, our optimization module takes as input the predicted runtime for a given EDA job on certain machine configuration (Section \ref{sec:prediction}), and the cost of running the job on a machine type recommended for that job (Section \ref{sec:char}).
In order to calculate the cost, we obtained the pricing table for the machine configurations from AWS at the time of this writeup, and calculated the total cost (in USD) for each EDA job (cost = runtime in hours $\times$ cost per hour).
To demonstrate our optimization, we applied different runtime constraints on predictions the of the \textit{sparc\_core} design as shown in Table \ref{tab:opt-results}.
Our algorithm outputs the recommended machine configuration for each job that minimizes the total cost subject to the given total runtime constraint. 
As we tighten the time constraint, we observe that the algorithm chooses higher machine configurations in some tasks (but not all).
A very tight time constraint cannot be met and no solution is presented.
Figure \ref{fig:cost} shows the cost savings that we get from running our optimization as compared to over-provisioning (using 8 vCPUs in all jobs) or under-provisioning (using 1 vCPU in all jobs) machines.
It offers an average of 35.29\% cost saving with minimal overhead to the best runtime.

\section{Conclusions}

We present an end-to-end workflow for a cost-efficient deployment of EDA workloads on the cloud.
Our method saves 35.29\% of the costs while meeting scheduled deadlines.
The code is open-source under a permissive license (BSD-3) and is available publicly on GitHub\footnote{https://github.com/scale-lab/EDAonCloud}.

\label{sec:conc}

\bibliographystyle{IEEEtran}

{\footnotesize
\bibliography{refbib}}

\begin{thebibliography}{1}
\providecommand{\url}[1]{#1}
\csname url@samestyle\endcsname
\providecommand{\newblock}{\relax}
\providecommand{\bibinfo}[2]{#2}
\providecommand{\BIBentrySTDinterwordspacing}{\spaceskip=0pt\relax}
\providecommand{\BIBentryALTinterwordstretchfactor}{4}
\providecommand{\BIBentryALTinterwordspacing}{\spaceskip=\fontdimen2\font plus
\BIBentryALTinterwordstretchfactor\fontdimen3\font minus
  \fontdimen4\font\relax}
\providecommand{\BIBforeignlanguage}[2]{{%
\expandafter\ifx\csname l@#1\endcsname\relax
\typeout{** WARNING: IEEEtran.bst: No hyphenation pattern has been}%
\typeout{** loaded for the language `#1'. Using the pattern for}%
\typeout{** the default language instead.}%
\else
\language=\csname l@#1\endcsname
\fi
#2}}
\providecommand{\BIBdecl}{\relax}
\BIBdecl

\bibitem{intel-cloud}
V.~{Kamath}, R.~{Giri}, and R.~{Muralidhar}, ``Experiences with a private
  enterprise cloud: Providing fault tolerance and high availability for
  interactive eda applications,'' in \emph{2013 IEEE Sixth International
  Conference on Cloud Computing}, 2013, pp. 770--777.

\bibitem{sehgal2016eda-ready-cloud}
N.~Sehgal, J.~M. Acken, and S.~Sohoni, ``Is the eda industry ready for cloud
  computing?'' \emph{IETE Technical Review}, vol.~33, no.~4, pp. 345--356,
  2016.

\bibitem{balkind2016openpiton}
J.~Balkind, M.~McKeown, Y.~Fu, T.~Nguyen, Y.~Zhou, A.~Lavrov, M.~Shahrad,
  A.~Fuchs, S.~Payne, X.~Liang \emph{et~al.}, ``Openpiton: An open source
  manycore research framework,'' \emph{ACM SIGPLAN Notices}, vol.~51, no.~4,
  pp. 217--232, 2016.

\bibitem{kipf2016semi}
T.~N. Kipf and M.~Welling, ``Semi-supervised classification with graph
  convolutional networks,'' \emph{arXiv preprint arXiv:1609.02907}, 2016.

\bibitem{Kellerer2004}
\BIBentryALTinterwordspacing
H.~Kellerer, U.~Pferschy, and D.~Pisinger, \emph{The Multiple-Choice Knapsack
  Problem}.\hskip 1em plus 0.5em minus 0.4em\relax Berlin, Heidelberg: Springer
  Berlin Heidelberg, 2004, pp. 317--347. [Online]. Available:
  \url{https://doi.org/10.1007/978-3-540-24777-7\_11}
\BIBentrySTDinterwordspacing

\bibitem{dudzinski1987exact}
K.~Dudzi{\'n}ski and S.~Walukiewicz, ``Exact methods for the knapsack problem
  and its generalizations,'' \emph{European Journal of Operational Research},
  vol.~28, no.~1, pp. 3--21, 1987.

\bibitem{amaru2015epfl}
L.~Amar{\'u}, P.-E. Gaillardon, and G.~De~Micheli, ``The epfl combinational
  benchmark suite,'' in \emph{Proceedings of the 24th International Workshop on
  Logic \& Synthesis (IWLS)}, no. CONF, 2015.

\bibitem{opencores}
\BIBentryALTinterwordspacing
``Opencores.'' [Online]. Available: \url{https://opencores.org/}
\BIBentrySTDinterwordspacing

\end{thebibliography}

\end{document}